# First experiments looking at electric field outside the Dense Plasma Focus show axial magnetic field exists before, during and after the pinch phase

Mladen Mitov, Alexander Blagoev, Stanislav Zapryanov and S K H Auluck

*Abstract*—Recent experiments using 15 frame interferometry on PF-1000 facility in Warsaw confirm the association between neutron emission and spontaneously self-organized, relatively long lasting, finite plasma structures. A crucial aspect of this association is the simultaneous observation of an axial magnetic field, which can allow magnetic flux lines to densely cover closed surfaces creating "magnetic flux surfaces". Evolution of such 3-dimensional (3-D) magnetic field structures is necessarily accompanied by induced electric field that can provide a very long (theoretically infinite) acceleration path length along a trajectory enclosed within the magnetic structure leading to high ion kinetic energy, resulting in a high reaction rate. Associated charge and current densities can be related to electric scalar potential and magnetic vector potential measured outside the plasma. We report our first observations of these fields outside the plasma focus and discuss their general features. The reported technique is capable of unambiguous first-principles interpretation of signals in terms of quantities related to distributions of charge density and rate of change of azimuthal current density ("electromagnetic structure") in the plasma focus. It is non-intrusive and completely insensitive to non-axisymmetric aspects of plasma. Our first results show that axial magnetic field generated by azimuthal current density distribution symmetric about the axis exists before, during and after the pinch phase.
.

## I. Introduction

RECENT experiments on PF-1000 [1-5] demonstrate association between the neutron production and some significant features of plasma dynamics observed using 15-frame interferometry. The Mach-Zahnder interferometer is set to have parallel mirrors. The resulting fringes are contours of equal linear electron density (integration of electron density over chords through the plasma along the path of the laser beam). Closed fringes therefore may be interpreted as representing bounded plasma structures under suitable assumptions[1]. Notable features of the observed plasma dynamics include relatively long-lived, spontaneously generated structures such as "plasmoids" (structures whose density has a single peak in the middle) and "toroids" (structures whose density has two bilateral peaks around the centre). These structures arise even before the arrival of the dense plasma sheath at the axis[1]. An important part is played in the interpretation of experiments by observations of an axial component of magnetic field using magnetic probes. Some of these observations pertain to the region ahead of the still-imploding dense plasma sheath, in a region where the toroidal structures are located[2]. Under appropriate conditions, the azimuthal magnetic field and the axial magnetic field can together make the magnetic flux lines densely cover closed surfaces, generating nested magnetic flux surfaces[6,7]. Since electron motion and electron thermal conduction are relatively unrestricted along the magnetic field (as compared with across the magnetic field), it is reasonable to expect that the magnetic flux surfaces should also be surfaces of equal electron density and equal electron temperature[5]. The bounded plasma structures revealed by closed fringes can therefore be conjectured to be associated with closed magnetic flux surfaces[5]. As the amplitude of magnetic field changes in time during evolution of these structures, the ions within should get accelerated by the induced electric field. The induced electric field shares topological features of the magnetic field, providing a long path length for acceleration which is folded within the finite volume of the bounded structure [7]. Thus, it is reasonable to expect that these plasma structures should be associated with neutron emission. This has indeed been confirmed[1]. The emission of the first, more isotropic pulse of neutrons, as well as the accompanying hard x-ray pulse, is observed to occur in coincidence with the formation and decay of bounded plasma structures.

Bounded long lived plasma structures have also been reported from the Nessi plasma focus in Stuttgart[8] using time-integrated soft x-ray photography and in the Frascati 1 MJ plasma focus using 3-frame schlieren[9]. The latter experiment also reported time correlation between the neutron emission and the plasma structure and a correlation between the physical location and size between the plasma structure and the neutron pinhole camera image[10]. The interpretation of bounded plasma structures in terms of closed magnetic surfaces has also been suggested earlier[11].

The association between the neutron emission and spontaneous self-organized bounded plasma structures is accompanied with the observation that the dominant energy of

This work has been partially supported by the contract 1827 of the Scientific Research Fund of the University of Sofia.

Mladen Mitov, Alexander Blagoev and Stanislav Zapryanov are with University of Sofia, Faculty of Physics, 5 J. Bourchierblvd., 1164, Sofia, Bulgaria.
e-mail addresses: Mladen Mitov (hann@abv.bg),
Alexander Blagoev(blagoev@phys.uni-sofia.bg),
Stanislav Zapryanov (sasho54@gmail.com)
S K H Auluckis with HiQTechKnowWorks Pvt. Ltd, Nerul, Navi Mumbai, India. He is currently Chairman, International Scientific Committee on Dense Magnetized Plasmas, Hery 23, P.O. Box 49, 00-908 Warsaw, Poland (skhauluck@gmail.com)

accelerated deuterons responsible for fusion neutrons is identical for neutron emission along the axial and radial directions[1] for the first, more isotropic neutron pulse that begins slightly before the stagnation of the dense plasma sheath on the axis. This accords well with the expectation that the acceleration of ions takes place along the electrical lines of force that densely cover a closed surface. This is supported by observations from multiple laboratories, using a variety of diagnostic techniques, that, when taken together, indicate that neutrons are produced by energetic ions moving in orbits whose turning points lie on a toroidal surface[7, 12].

It might be reasonably inferred from these mutually supportive experimental results that the plasma has a 3-dimensional electromagnetic structure consisting of time and space dependent distributions of charge density and current density. The charge density arises from charge separation over spatial scales comparable with the Debye length. It can be calculated by taking divergence of the Generalized Ohm's Law. Because of the curved shape of the plasma in all phases of evolution, the term $\partial_z(-v_r B_\theta)$ is always non-zero ensuring presence of charge separation on the plasma. The charge density distribution acts as a source of electric scalar potential throughout space, especially in regions just outside the plasma. Particularly important is the role of the azimuthal component of current density. Kubes et al [4] mention that "the evolution of plasma structures starts at an increase in the self-organized azimuthal current component....". It is shown below that the azimuthal current density produces an azimuthal component of magnetic vector potential outside the plasma that can be directly measured. This allows a new kind of diagnostics to be developed for DPF experiments that is complementary to already fielded diagnostics.

This paper reports the first experiments aimed at measurement of electric scalar potential and azimuthal component of magnetic vector potential outside the dense plasma focus. We emphasize that *this paper does not aim to measure local axial magnetic field* inside the plasma, which would require insertion of appropriate magnetic probes into the plasma. Rather, this paper aims to address the question: does azimuthal current exist in the plasma whose Fourier development in angular coordinates has an angle-independent term? Our preliminary data answers the question in the affirmative, allowing us to make a definitive statement concerning existence (but not quantitative measurement) of axial magnetic field during evolution of the plasma focus.

The next Section II establishes the theoretical background of the measurement, Section III describes the experimental set up and Section IV the results. Section V discusses possible sources of error and argues that while there is certainly scope for improvement, our main conclusions remain unaffected. Section VI concludes the paper with a brief summary and remarks.

## II. THEORY OF MEASUREMENT UNDERLYING THE NEW DIAGNOSTIC

The interferometric results on PF-1000[1-5] prove that spatially non-uniform plasma structures appear, evolve and decay within a time period of a few hundred nanoseconds. Laplacian of the associated plasma potential should be proportional to the charge density distribution in the plasma. The inference of a time-varying poloidal magnetic field from magnetic probe measurements indicate the existence of a toroidal current density.

Charge and current density distributions act[13] as sources of electric scalar potential $\varphi(\vec{r},t)$ and magnetic vector potential $\vec{A}(\vec{r},t)$, which are related to the electric and magnetic field by the relations

$$\vec{B} = \vec{\nabla} \times \vec{A} \ , \ \vec{E} = -\vec{\nabla}\varphi - \partial_t \vec{A} \ , \ \vec{\nabla}\cdot\vec{A} + \mu_0\varepsilon_0\partial_t\varphi = 0 \qquad 1$$

*throughout space – even outside the plasma* as

$$\varphi(\vec{r},t) = \frac{1}{4\pi\varepsilon_0}\int d^3\vec{r}' \frac{\rho(\vec{r}',t')}{|\vec{r}-\vec{r}'|} \qquad 2$$

$$\vec{A}(\vec{r},t) = \frac{\mu_0}{4\pi}\int d^3\vec{r}' \frac{\vec{J}(\vec{r}',t')}{|\vec{r}-\vec{r}'|} \qquad 3$$

$$t' = t - |\vec{r}-\vec{r}'|/c \qquad 4$$

Consider now a conductor placed outside the cathode in the shape of an open cylindrical ring of thickness d, inner radius $R_1$ and outer radius $R_1 + \Delta R$, centered on the axis of the device at an elevation $z_0$ along the z axis, with a coaxial cable connected to its ends (along with suitable impedance), which is taken to an oscilloscope and is terminated in its characteristic impedance. In the case of plasma focus, the delay term in (4) can be neglected on the time scale of plasma dynamics.

The potentials at an arbitrary point $(R,\theta,Z)$ on this ring at time t would then be given by multipole expansions[13], with $R_0 \equiv \sqrt{R_1^2 + z_0^2}$

$$\varphi(R,\theta,Z,t) = \frac{1}{4\pi\varepsilon_0}\sum_{n=0}^{\infty}\frac{1}{R_0^{n+1}}$$
$$\times \int_0^{R_b} r'dr' \int_0^{2\pi} d\theta' \int_0^{Z_b} dz' \rho(R',\theta',Z',t) P_n(\cos(\vec{r},\vec{r}'))|\vec{r}'|^n \qquad 5$$

$$\vec{A}(R,\theta,Z,t) = \frac{\mu_0}{4\pi}\sum_{n=0}^{\infty}\frac{1}{R_0^{n+1}}$$
$$\times \int_0^{R_b} r'dr' \int_0^{2\pi} d\theta' \int_0^{h_b} dz' \vec{J}(R',\theta',Z',t) P_n(\cos(\vec{r},\vec{r}'))|\vec{r}'|^n \qquad 6$$

Since the ring is conducting, all points on it reach the same scalar potential, which can be estimated as the volume average of the scalar potential given by (6):

$$\varphi_{ring}(t) = \frac{1}{2\pi R_1 \Delta R d}\int_{R_1}^{R_1+\Delta R} RdR \int_0^{2\pi-\delta} d\theta \int_{z_0}^{z_0+d} dz\, \varphi(R,\theta,Z,t)$$
$$= \frac{1}{2\pi}\int_0^{2\pi-\delta} d\theta\, \varphi(R_1,\theta,z_0,t) \qquad 7$$

where $\delta = g/R_1 \ll 2\pi$ and g is the gap in the open ring.

On the other hand, *the two ends of the open ring should have*

*a difference in potential equal to the line integral of the electric field* $\oint \vec{E} \cdot d\vec{\ell}$ *along the ring*:

$$\chi(t) = \oint \vec{E} \cdot d\vec{\ell} \approx \int_0^{2\pi-\delta} R_1 d\theta \left(\hat{\theta} \cdot \vec{E}(R_1,\theta,z_0,t)\right)$$

$$= R_1 \int_0^{2\pi-\delta} d\theta \left(-\partial_\theta \varphi - \frac{\partial A_\theta}{\partial t}\right)$$

$$= R_1 \delta \cdot \varphi(R_1,0,z_0,t) - R_1 \frac{\mu_0}{4\pi} \sum_{n=0}^\infty \frac{1}{R_0^{n+1}} \int_0^{2\pi-\delta} d\theta \int_0^{2\pi} d\theta' \quad 8$$

$$\times \int_0^{R_b} R'dr' \int_0^{h_b} dZ' P_n\left(\cos(\vec{r},\vec{r}')\right)|\vec{r}'|^n \frac{\partial}{\partial t} J_\theta(R',\theta',Z',t)$$

The physical meaning of the two expressions (7) and (8) needs to be properly appreciated. Expression (7) is the volume average of a scalar, periodic function of $\theta$, that can be expressed in terms of a Fourier series in $\theta$. Integration over the range $[0,2\pi]$ would have contribution *only from the Fourier series term independent of periodic functions of $\theta$*. Moreover, the volume average would remain invariant under the transformation $\theta \to -\theta$. In the case of integration over the range $[0, 2\pi - \delta]$, the invariance under the transformation $\theta \to -\theta$ is not affected; the contribution of periodic terms would be proportional to $\delta$ and can be made negligible for $\delta \ll 2\pi$, which can be physically realized.

The expression (8), on the other hand, is line integral of a vector quantity that reverses its sign under the transformation $\theta \to -\theta$. The integral over $\theta$ retains contribution only from the $\theta$-independent term in Fourier series of the integrand; the contribution from the periodic terms remains proportional to $\delta$. In the limit of small $\delta$, it can be asserted that the quantity $\chi(t)$ represents time variation of the *$\theta$-symmetric portion* of the rate of change of the azimuthal component of vector potential.

The net potential seen by the two ends of the open ring are $\varphi_{ring}(t)$ and $\varphi_{ring}(t) + \chi(t)$. When connected to the differential amplifier input of the oscilloscope using a cable, the quantity $\varphi_{ring}(t)$ appears as a common mode signal. The differential amplifier of the oscilloscope will produce an output given by

$$V_{Signal} = A_d \chi(t) + A_{CM} \varphi_{ring}(t) \qquad 9$$

where $A_d$ is the differential mode gain of the input amplifier and $A_{CM}$ is its common mode gain. The oscilloscope input differential amplifier specifications describe a Common Mode Rejection Ratio (CMRR) as

$$CMRR(dB) = 20 Log_{10}\left(\frac{A_d}{A_{CM}}\right) \qquad 10$$

so that

$$A_{CM} = A_d 10^{-\left(\frac{CMRR(dB)}{20}\right)} \qquad 11$$

For a pair of identical rings connected clockwise (CW) and counterclockwise (CCW) the sign of $\chi(t)$ is reversed. Defining

$$DIFF = (V_{CCW} - V_{CW})/2 ; SUM = (V_{CCW} + V_{CW})/2; \qquad 12$$

it is clear that

$$DIFF = A_d \chi(t); SUM = A_{CM} \varphi_{ring}(t) \qquad 13$$

The above theoretical analysis suggests that it should be possible, in principle, to deploy many vertically displaced stacks of concentric, coplanar pairs of rings of different radii, connected in CW and CCW sense, with corresponding measurements of $\varphi_{ring}(t)$ and $\chi(t)$. One could then determine the time variation of contributions from various terms of the multipole expansion of charge density and current density, truncated to the number of terms equal to the number of ring pairs, for various elevations leading to 3-dimensional reconstruction of the *$\theta$-symmetric parts* of azimuthal current density and charge density. It should also be possible to position many floating conductors (say, discs) around the plasma, all connected to oscilloscope channels, allowing a tomographic determination of the *$\theta$-dependent* charge density distribution, resulting from discrete self-organized plasma structures that could have complex 3-dimensional shapes, such as those arising from generation and destruction of filaments and off-axis point plasmas [8].

This discussion shows that the proposed diagnostic
1. has an unambiguous first-principles theoretical interpretation,
2. is completely insensitive to asymmetric phenomena such as restrikes, or m≠0 unstable modes or any other non-axisymmetric aspects of plasma formations *because of the double integration over angular coordinates of both source points and field points*,
3. does not perturb the plasma in any way.
4. and has the potential for increased sophistication that can lead to a 3-dimensional reconstruction of the electromagnetic structure of the DPF.

The purpose of the present paper is to provide a proof-of-concept of this idea before attempting an exhaustive investigation of plasma focus physics.

### III. EXPERIMENTAL SETUP

This section provides a short description of the Mather type PF device at Sofia University. The condenser bank consists of four 5 µF capacitors (the bank total capacitance is 20 µF) with maximal charging voltage of 40 kV. The main switch of the bank is a vacuum spark gap. The anode is a hollow copper tube with 2 cm diameter and 14.5 cm length, the cathode consists of six copper rods (0.8 cm diameter, 16 cm length) mounted in the massive cathode bottom on a circle with a 3.5 cm radius. The insulator is a quartz tube (2.6 cm in diameter, 3 cm length in the chamber). The stainless steel vacuum chamber has 15.5 cm inner diameter and 35 cm height.

The working gas can be air, Ar or Deuterium. The voltage operating range in the case of air or Ar is 15-18 kV and the nominal pressure is in the range of 0.8-2.0 mBar. Thus the stored energy of the PF does not exceed 3 kJ.

We are monitoring the discharge current, (Rogowski belt), the current derivative (pick-up coil), the soft X-ray (PIN diodes)

and the hard X-ray emission (scintillator probe) from the plasma. Thermo-luminescent dosimeters (TLD) are used to measure the full X-ray dose in the chamber of the PF and to control the radiation outside the device. The data of all of the detectors mentioned above have been recorded by two 4 channel Tektronix oscilloscopes (TDS 3034C, TDS 2004B).

We report results from two series of experiments.

*A. Physical construction and electrical connection*

The first exploratory series used single turn coils made of 2 mm thick copper wire supported on a 11.5 cm diameter bakelite tube. The gap in the two ends of each ring was significant and coaxial cables were soldered to the coil ends as shown in Fig. 1

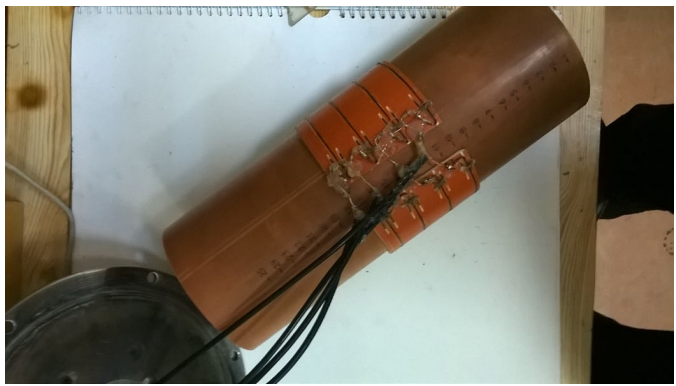

Fig 1: Single turn coils used in the first series of experiments. The coaxial cables are soldered across the wire loops but there is a large azimuthal gap. The brown colored tube is made of bakelite

The second series deployed a pair of rings of copper, of 15 micron thickness, 110 mm and 130 mm inner diameters (labeled rings 1 and 2) and 2 mm width, machined on two printed circuit board (PCB) laminates (labeled A and B) of 1.6 mm thickness. Each ring was cut creating a circumferential gap of about 0.5 mm (see fig 2). The rings were placed on top of each other, so that both copper layers faced towards the base of the cathode. The spacing between the copper rings was therefore 1.6 mm.

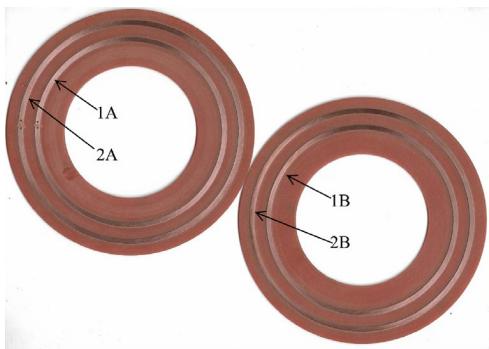

Fig 2: Rings of copper machined on a printed circuit board (PCB) phenolic laminate used in Series 2 of experiments. The rings have internal diameters of 110 and 130 mm and annular width of 2 mm. The inner rings are used for the present measurements

Electrical connections were made with ~30 cm long color-coded twisted insulated wires soldered across the gap and connected via a vacuum feed-through to an external coaxial cable. Metallic wire-braid screening was provided over the twisted pair of wires up to the vacuum feed-through. This wire-braid was in turn covered with a plastic insulation tape to avoid accidental ground loops. A 1/4 watt 50 Ohm resistance was soldered in series with the ring.

In both series of experiments, the two ends of the single turn coil were spatially quite close and electrically connected with a solid conductor of less than milli-Ohm resistance. The 50 Ohm resistance in series with the coil ensured that current flow in the circuit produced a negligible voltage drop around the loop and the voltage was measured under effectively open-circuit conditions. The only difference between the two ends of the coil was the azimuthal angle $\theta$ of the ring conductor connecting them: for one end it was close to zero and for the other, it was close to $2\pi$. The electrical connections were made in such manner that in case of ring 1A, the center conductor of the coaxial cable was electrically connected to the end of the ring close to zero azimuthal angle and the cable shield was connected to the other end. The ring 1B, having the same radius but axially displaced by 1.6 mm, was connected in the reverse azimuthal sense. A schematic of electrical connection is shown in Fig 3.

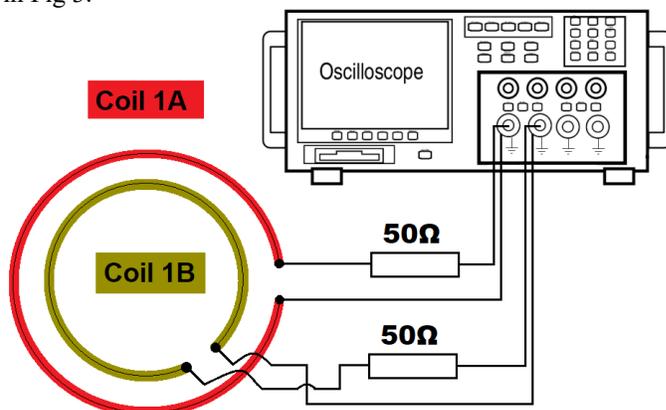

Fig 3: Schematic of electrical connections

*B. Physical location*

In the first series, the plastic tube supporting the single turn coils rested on the cathode flange in such manner that the expected region of the pinch was approximately at the centre of two of the coils. The scheme of placement for the first series is shown in Fig 4.

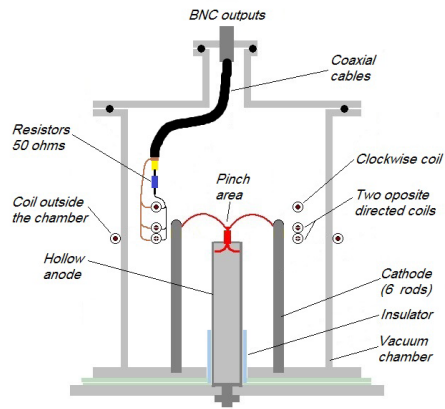

Fig. 4: Schematic of placement of the coils in the first series of measurements

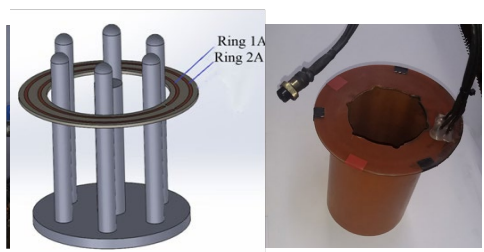

Fig 5a          Fig5b          Fig 5c

Fig 5. Placement of the PCB laminate supporting the rings. (a) Only one laminate is shown facing up for clarity in the schematic, which illustrates the position of the rings with respect to the anode and cathode. Physically, the copper rings are facing down towards the cathode base (b) Physical support for the rings. The electrical connection shown consists of two pairs of twisted wires covered with a metallic braided sleeve covered with electrical insulating tape to prevent accidental ground loops (c) View through the port

In the second series, it was possible to place two identical rings in close proximity, axially separated by 1.6 mm, the thickness of the PCB laminate. The physical location of the rings in the plasma focus is illustrated in Fig 5. The rings were protected from the plasma by a layer of opaque plastic glued to the PCB. A bakelite tube was used to support the rings in such manner that the rings were outside the cathode and at the elevation of the pinch zone.

It is clear that in both cases, the rings were well outside the cathode so that the flux of the azimuthal magnetic field could not have any linkage with the measurement circuit formed by the rings

## IV. RESULTS

The purpose of this report is to illustrate the viability and significance of the proposed new diagnostic. This is intended as a proof-of-concept: we do not intend to present a systematic study of plasma focus phenomenology using this diagnostic at the moment, which involves a more extensive preparation.

The results for the first series of experiments [14] show that signals for CW and CCW coils have a noticeable difference. The SUM and DIFF waveforms exhibit sharp phase-transition-like features correlated with extrema of current derivative that persist throughout the duration of the discharge. However, this series of shots had a significant experimental flaw in the form of a large gap between the ends of the coil, so that at this stage, a credible interpretation of reported waveforms[14] is not feasible.

This flaw was corrected in the second series of experiments where the clockwise and anti-clockwise connected loops are machined on a printed circuit board (see Fig 2), ensuring its planarity and conforming closely to the assumptions of section II.

Fig. 6 shows an example where the signals corresponding to the clockwise and anticlockwise connected loops are significantly different.

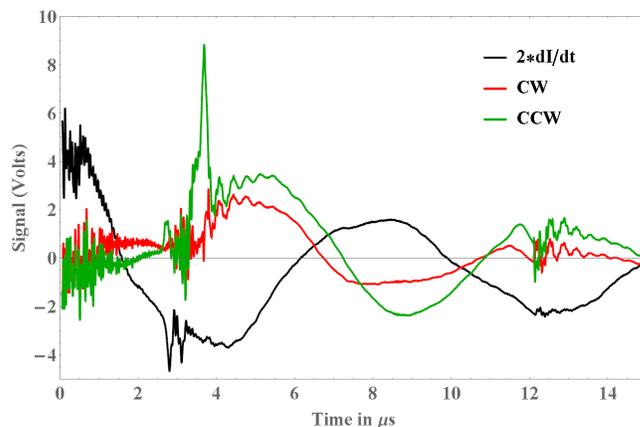

Fig 6. The clockwise-connected (CW) and counter-clockwise-connected (CCW) machined loops shown in Fig 2 and placed in close proximity produce signals which are strikingly different. Data from Shot #519 (p = 1.13 Torr Ar, 17.0 kV, 10th May, 2019), 30 point moving-average smoothening. The dI/dt signal and the loop signals have clearly different shapes with features which are time-correlated

This difference is difficult to interpret in any manner other than the resultant of two signals, one of which is symmetric in the azimuthal angle and the other one is anti-symmetric. Fig. 7 shows the corresponding SUM and DIFF signals.

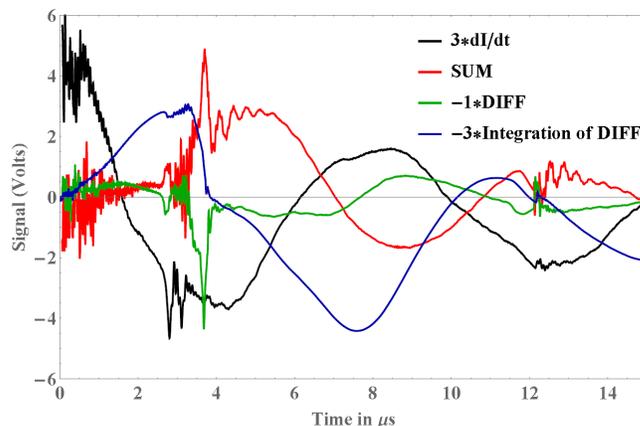

Fig 7. The SUM and DIFF signals corresponding to the signals for Fig 8. Integration of DIFF takes into account base-line correction calculated from the last microsecond of the 40 microsecond record. The DIFF signal is multiplied by -1 for clarity of display in order to avoid overlap with the SUM signal.

According to the discussion of Section II, the DIFF signal should depend on the time derivative of azimuthal current density. Fig 7 therefore also displays the integration of the DIFF signal, which should represent the azimuthal current density. The integration-of-DIFF curve in Fig 7 rises on a time scale comparable with the discharge quarter cycle time followed by a sharp dip coinciding with the first dI/dt dip. There is however, a second dI/dt dip which is followed by a much sharper and deeper dip in the DIFF signal. The integration of the DIFF signal falls over a time short compared with the quarter cycle time and reverses its sign. It has some activity around 12 microseconds, coinciding with activity in the dI/dt signal. This data gives a strong indication that the signals induced in the loops, placed outside the cathode and hence decoupled from azimuthal magnetic field of the main discharge current, provide significant physical information about phenomena taking place within the plasma in a completely non-perturbative manner. *It is especially important to note that this diagnostic is not sensitive to non-symmetric aspects of plasma behavior*.

The slowly rising initial portion of the integration-of-DIFF curve in Fig 7 corresponding to the rundown phase is consistent with the discussion of role of the geomagnetic field by Mather and Ahluwalia [15]. The first sharp dip of the DIFF signal coincides with the first current derivative dip but the second larger sharp dip of DIFF signal is not correlated with any feature of the current derivative signal. *This constitutes clear proof that the signal generated by the new diagnostic is a new physical measurement independent of the main discharge current or of plasma leaking between the cathode bars or any parasitic coupling between the electrodes and the coils or ground loops (see Section V)*. The integration of the DIFF signal persists for a long time after the current derivative singularity. This is consistent with observations of compact plasma objects long after the pinch phase [16,17], which are conjectured to be laboratory analogs of astrophysical jets.

First results from the new diagnostic described in this paper therefore show clear evidence of presence of axially symmetric azimuthal current and associated axial magnetic field before, during and after the pinch phase.

## V. Potential sources of error and their impact

The purpose and scope of this paper has already been mentioned above but is repeated here to provide the context for the discussion of potential errors that follows.

The purpose is to propose and demonstrate the viability and significance of a new diagnostic not tried so far in the case of plasma focus. Perfection of the proposed idea is an agenda for the future, not covered in the scope of the present paper. Hence it is relevant to discuss potential sources of error, their impact on the conclusions of this report and possible mitigation strategies for future.

As already emphasized above, *this diagnostic does not aim to measure the local axial magnetic field in the plasma - a common misconception among those coming across this technique for the first time*. It specifically limits itself to measuring the line-integral of time-derivative of azimuthal component of magnetic vector around a circular loop - a well-defined experimental number. Discussion of *experimental errors* must therefore be restricted to the quantity being measured, which is neither the local axial magnetic field nor the voltage on the pinch nor the common mode signal.

The *interpretation* of the measured quantity in terms of the angle-independent term of the angular Fourier representation of the azimuthal component of the time-derivative of the magnetic vector potential at the location of the diagnostic is a major takeaway of this paper. Experimental errors that affect this interpretation and its downstream consequences (elaborated below) are therefore likely to have a significant impact on the viability and significance of this diagnostic. For example, experimental proof of *existence* of the angle-independent term of the angular Fourier representation of the azimuthal component of the time-derivative of the magnetic vector potential at the location of the diagnostic *strongly implies* existence of azimuthal current having a non-zero angle-independent term in its angular Fourier representation. This, in turn implies *existence* of a local axial magnetic field within the plasma. Although its quantitative estimate is not obtained from the diagnostic, proof of its existence itself is of great significance since axial magnetic field is neither predicted nor required by most theoretical approaches to the plasma focus.

This section retrospectively looks at the potential sources of experimental error, their impact on the reported findings and possible mitigation strategies for the future.

1) "Parasitic" coupling between the electrode system and the coils: This coupling is an electrostatic effect and cannot distinguish between CW and CCW coils. Hence it cannot affect the DIFF signal, which is used in the interpretation of the experiment.
2) Coupling of high frequency noise to the bare coil: This produces a noisy signal with high frequency oscillation over-riding a structure that has temporal features correlated with the current derivative signal. In the present experiment, this has been handled through the moving-average smoothing technique, choosing the number of neighboring points that best reproduces the general shape of the dI/dt, CW and CCW signals.
3) Ground loops: Since the signal has an expected common-mode component, the common-mode gain of the input amplifier of the oscilloscope comes into picture. Ground loops, caused by multiple-point grounding of the measurement and pulse-power circuits, can cause noise signals in the common mode circuits. High current driven by the pulse power discharge into the ground loop circuits can additionally introduce differential mode error signal through mirror current (mutual inductance) coupling within the coaxial cables. *Such differential error signals, however, cannot distinguish between CW and CCW signals*. Ground loop coupling should therefore induce identical error signals in both the CW and CCW measurement channels on the same oscilloscope so that the DIFF signal is not affected. In particular, ground loops cannot cause CW signals to be different from CCW signals. As emphasized in Section IV, the striking difference between the observed double-dip structures of the DIFF signal and the current-derivative signal cannot be caused by ground loop coupling. This observation, however, is not intended to minimize the undesirability of ground loops. The best way to avoid ground loops is to operate the oscilloscope on a battery-based invertor completely isolated from power mains, power ground and insulated from all nearby metallic objects. This however has its own associated risks and costs.

4) Errors from misalignment of plasma and the coils: The coils and the bakelite tube support are machined components that are snug fitted together with the squirrel cage cathode so that centering and angular errors with respect to the electrode structure are comparable with general manufacturing tolerances of the plasma focus device. Errors under this category can arise only from the inherent fluctuations in the position of the plasma focus with respect to the electrode axis which is an inherent part of the plasma focus phenomenon. Essentially, this error amounts to a rotation and/or a translation between coordinate systems attached with the plasma and the coils. This cannot affect conclusions concerning the existence of the quantity being measured, which is all that is being claimed in the present paper.

## VI. Summary and conclusions

In this paper, we introduce and provide experimental proof-of-concept of a new kind of diagnostic. This technique records the time variation of the rate of change of azimuthal component of magnetic vector potential, generated outside the plasma, by azimuthal component of plasma current density that is symmetric about the axis. It also provides the time variation of azimuthally symmetric part of electric scalar potential, generated outside the plasma by charge separation associated with plasma evolution. This diagnostic

1. has an unambiguous first-principles theoretical interpretation,
2. is completely insensitive to asymmetric phenomena such as restrikes, or m≠0 unstable modes or any other non-axisymmetric aspects of plasma formations,
3. does not have any parasitic contribution from the electric or magnetic field associated with the main discharge current
4. is not affected by plasma leaking between cathode bars
5. does not perturb the plasma in any way.
6. and has the potential for increased sophistication that can lead to a 3-dimensional reconstruction of the electromagnetic structure of the DPF.

Its significance is that it can provide unambiguous information about the instant of generation and duration of inhomogeneous plasma structures that have an azimuthal component of current density which is necessary for formation of magnetic flux surfaces.

Our first results show that axial magnetic field generated by axially symmetric azimuthal current exists before, during and after the pinch phase. While more investigations are needed to interpret these signals, our preliminary impression is that these signatures are associated with spontaneous formation of bounded plasma structures having bounded magnetic flux surfaces, which have been observed previously using imaging techniques [1-5,8,9,10].

This diagnostic has much scope for improvement. The presence of a common mode signal is an undesirable feature that could be affected by ground loops, perhaps requiring floating battery operated oscilloscopes. The self-inductance of the ring creates impedance mismatch problems with the coaxial cable, which would be unmanageable for large devices. This may require development of a distributed capacitance version of the diagnostic. It would also be desirable to perform in situ absolute calibration of the diagnostic in terms of total azimuthal current and voltage on the axis.

**Acknowledgements**:
This work has been partially supported by the contract 1827 of the Scientific Research Fund of the University of Sofia. Useful suggestions of H. Bruzzone are gratefully acknowledged.

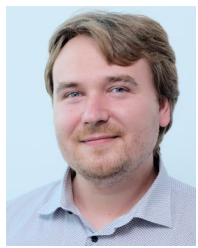
**Mladen Boykov Mitov,** graduated from the Technical University of Sofia with MS in 2011, PhD in2016. In the list of publications - 4 Papers in research journals and twelve reports in scientific conferences in the field of application of electrical probes in low and high temperature plasma. He is now a student in the Master's program ofthe Faculty of Physics at the University of Sofia.

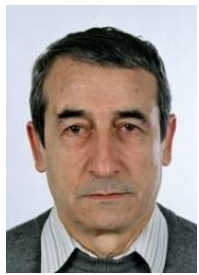
**Alexander Borissov Blagoev**.graduated as MS in the Faculty of Physics at the University of Sofia, obtained his PhD in 1973 in the Leningrad State University,  D Sc in 1997 in the Faculty of Physics at Sofia University. In 2009, he retired as a Professor in the Faculty of Physics. Since then, he is with Scientific Research Center of Sofia University. His publication list comprises 90 papers and reports on scientific conferences in the field of low and high temperature plasma

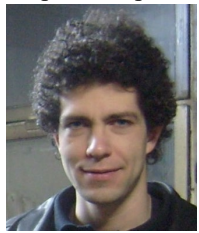
**Stanislav Hristov Zapryanov**, Graduated in the Faculty of Physics at the University of Sofia, MS in 2011, PhD in 2014. He has 5 Papers in research journals and several reports on scientific conferencesin the field of Plasma Focus research.

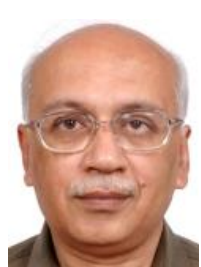
**Sunil Kumar Hiralal Auluck** did his M.Sc. in Physics from the Indian Institute of Technology, Kharagpur in 1977, and obtained Ph.D degree from University of Bombay in 1988. He joined Bhabha Atomic Research Center, Mumbai, India in 1978 and retired as Outstanding Scientist and Professor in the Homi Bhabha National Institute, a Deemed University, in 2014. He has been representing India on the International Scientific Committee on Dense Magnetized Plasmas (http://www.icdmp.pl/isc-dmp) since 2004 and is currently its Chairman.